\documentclass{article}

\usepackage{url, setspace, times, amsmath, amssymb, latexsym, epsfig, subfigure}

\long\def\remove#1{}

\def\RR{\mathbb R}

\newsavebox{\smallProofsym}                            

\savebox{\smallProofsym}                               %
{
\begin{picture}(6,6)                                   %
\put(0,0){\framebox(6,6){}}                            %
\put(0,2){\framebox(4,4){}}                            %
\end{picture}                                          %
}

\newcommand{\qed}{\hfill$\square$\bigskip}

\newcommand{\proof}{\noindent {\bf Proof}.~~}

\newcommand{\hide}[1]{}

\newtheorem{definition}{Definition}[section]
\newtheorem{theorem}{Theorem}[section]
\newtheorem{lemma}{Lemma}[section]

\begin{document}

\title{On Locally Gabriel Geometric Graphs}
\date{}

\newcommand{\atgen}{\symbol{'100}}

\author{
Sathish Govindarajan \thanks{
Department of Computer Science and Automation,
Indian Institute of Science, Bangalore, India.
E-mail: \texttt{gsat\atgen{}csa.iisc.ernet.in}
}
\and
Abhijeet Khopkar \thanks{
Department of Computer Science and Automation,
Indian Institute of Science, Bangalore, India.
E-mail: \texttt{abhijit\atgen{}csa.iisc.ernet.in}
}
}

\maketitle
\thispagestyle{empty}

\begin{abstract}
Let $P$ be a set of $n$ points in the plane. A geometric graph $G$ on $P$ is said to
be {\it locally Gabriel} if for every edge $(u,v)$ in $G$, the disk with $u$ and $v$
as diameter does not contain any points of $P$ that are neighbors of $u$ or $v$ in $G$.
A locally Gabriel graph is a generalization of Gabriel graph and is motivated by
applications in wireless networks. Unlike a Gabriel graph, there is no unique locally Gabriel graph 
on a given point set since no edge in a locally Gabriel graph is necessarily included or excluded.
Thus the edge set of the graph can be customized to optimize certain network parameters depending on the application.
In this paper, we show the following combinatorial bounds on edge complexity and independent sets of locally Gabriel graphs:

\begin{enumerate}
\item[(i)] For any $n$, there exists locally Gabriel graphs with $\Omega(n^{5/4})$ edges.
This improves upon the previous best bound of $\Omega(n^{1+\frac{1}{\log \log n}})$.

\item[(ii)] For various subclasses of convex point sets, we show tight linear bounds
on the maximum edge complexity of locally Gabriel graphs.

\item[(iii)] For any locally Gabriel graph on any $n$ point set, there exists an independent set of
size $\Omega(\sqrt{n}\log n)$.
\end{enumerate}
\end{abstract}


\section{Introduction}
A geometric graph $G=(V,E)$ is an embedding of the set $V$ as points in the plane and edges in $E$ as straight-line segments connecting the points in $V$.
Delaunay graphs, Gabriel graphs and Relative Neighborhood graphs (RNG) are fundamental
geometric proximity graphs with applications in fields like computer graphics, vision,
GIS, wireless networks, etc. For a nice survey on these graphs and their applications, see~\cite{JT}.

The Gabriel graph introduced by Gabriel and Sokal \cite{GS} is defined as follows: Given a set of points $P$ in the plane, an edge exists between points $u$ and $v$ iff
the Euclidean disk with $u$ and $v$ as diameter does not contain any other point of $P$.
Gabriel graphs have been used to model the topology in wireless networks~\cite{Bose,Urut}.
Motivated by applications in wireless networks, \cite{LCW02,KL10} generalized these structures
to $k$-locally delaunay/gabriel graphs. The edge complexity of these structures have been studied in \cite{KL10,PS04}.
In this paper, we focus on 1-locally Gabriel graphs and call them as {\em Locally Gabriel Graphs} ($LGG$s).

A {\it locally gabriel graph} is a geometric graph $G$ with the following property: for each edge $(u,v)$ in $G$,
the Euclidean disk with $u$ and $v$ as diameter does not contain any points of $P$ that are neighbors of $u$ or $v$ in $G$. 

Study of these graphs was initially motivated by design of dynamic routing protocols for \emph {ad hoc} wireless networks \cite{Li}.
An ad-hoc wireless network consists of a collection of wireless transceivers (corresponds to points) and an
underlying network topology (corresponds to edges) that is used for communication/routing.
Like Gabriel Graphs, $LGGs$ can be used to design wireless network topology since they capture the interference patterns well.
An interesting point to be noted is that there is no unique $LGG$ on a given point set since no edge in $LGG$ is necessarily included or excluded.
Thus the edge set of the graph (used for wireless communication) can be customized to optimize certain network parameters depending on the application.
$LGGs$ also provide certain advantages over Gabriel Graphs. While a Gabriel graph has linear number of edges (planar graph), we show in this paper that there exists $LGG$s with $n^{5/4}$ edges. A dense network can be desirable for applications like
broadcasting or multicasting where a large number of pairs of nodes need to communicate with each other.
Another important parameter in the topology of wireless network is the number of simultaneous transmissions that can be performed. A node in a wireless network cannot transmit and receive in the same time slot. Thus, the set of transmitting nodes at any time slot form an independent set in the underlying graph. We show that there exists an independent set of size $\Omega(\sqrt{n}\log n)$ in any $LGG$ of any $n$ pointset.

An interesting combinatorial question, that we address in this paper, is to bound the edge complexity of locally gabriel graphs.

It was observed in~\cite{PS04} that the unit distance graph~\cite{Erdos}, introduced by Erdos, is also a locally delaunay/gabriel graph.
The maximum edge complexity of unit distance graphs has been extensively studied~\cite{Erdos,Szek,ST}. See~\cite{BP04} for a survey
on this problem. There is a significant gap between
the lower and upper bounds and improving them is considered a hard open problem in discrete geometry. The edge complexity of unit distance graphs
on convex point sets have also been studied. The best lower bound is $2n-7$~\cite{EH91} and the best upper bound is $n \log n$~\cite{F90,BP01}.
It has been conjectured in~\cite{BP04} that the edge complexity of unit distance graphs on convex point sets is $2n$.

~\cite{KL10} initiated the study of maximum edge complexity of locally delaunay/gabriel graphs
by showing non-trivial upper bounds. ~\cite{PS04} showed an upper bound of $O(n^{3/2})$ and a
lower bound of $\Omega(n^{4/3})$ on the maximum edge complexity of locally delaunay graphs.

For locally gabriel graphs,~\cite{KL10} showed an upper bound of $O(n^{3/2})$
by proving that $K_{2,3}$ is a forbidden subgraph. The best known lower
bound is $\Omega(n^{1+\frac{1}{\log \log n}})$~\cite{Erdos}, given by
Erdos classic lower bound construction for unit distance graphs.
While the gap between the upper and lower bounds for locally delaunay graphs
has been narrowed significantly, the gap is quite wide for locally gabriel graphs.
In this paper, we improve the lower bound significantly.

We show the following results in this paper:
\begin{enumerate}
\item[(i)] For any $n$, there exists locally gabriel graphs with $\Omega(n^{5/4})$ edges.
This improves the previous lower bound of $\Omega(n^{1+\frac{1}{\log \log n}})$~\cite{Erdos}.
\item[(ii)] For various subclasses of convex point sets like monotonic convex point set,
half convex point set, centrally symmetric convex point set, we prove tight linear bounds on the
edge complexity of locally gabriel graphs.
\item[(iii)] For any $LGG$ on any $n$ point set, we show that there exists an independent set of
size $\Omega(\sqrt{n}\log n)$.
\end{enumerate}


The paper is organized as follows: Definitions that will be used in the paper is presented
in Section 2. We present the lower bound construction in Section 3 and analyze it in
Section 4. We prove various upper and lower bounds for convex point sets in Section 5. The independent set construction
is presented in Section 6.

\section{Preliminaries}
Let $P$ be a set of $n$ points in $\RR^2$. For any $p,q \in P$, we denote by $d_{pq}$ the disk with $p$ and $q$ as diameter.
\begin{definition}\emph{(Locally Gabriel condition)} Let $G_P$ be a geometric graph on $P$. An edge $(u,v)$ of $G_P$
is said to satisfy the locally Gabriel condition if disk $d_{uv}$ does not contain neighbors of $u$ or $v$ in $G_P$.
\end{definition}
\begin{definition}\emph{(Locally Gabriel Graph)}
A geometric graph $G_P$ on $P$ is said to be Locally Gabriel Graph (LGG) if every edge of $G_P$
satisfies the locally Gabriel condition.
\end{definition}
Let $p=(p^x,p^y)$ be any point in $\RR^2$.
\begin{definition}\emph{(Upper-right monotonic convex point set)}
Let $P = \{p_1,p_2\dots,p_k\}$ be a set of points in convex position that are
ordered in counterclockwise direction. $P$ is called a upper-right monotonic convex point set
if $p_i^x \leq p_j^x, p_i^y \geq p_j^y, \forall 1\leq i < j \leq k$
\end{definition}

\begin{figure}[htp]
\centering
\includegraphics[height=2.5cm]{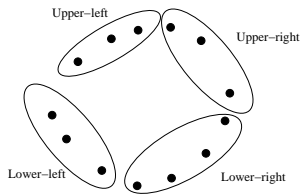}
\caption{Four types of monotonic convex sets}\label{fig:monotone}
\end{figure}

Similarly, we define the other three types of monotonic convex point sets, i.e.,
upper-left, lower-right and lower-left. Figure~\ref{fig:monotone} shows the 4 types of monotonic
convex point sets. Note that any convex point set can be decomposed into the above 4 types.
\begin{definition}\emph{(Half convex point set)}
Let $P=Q\cup R$ be a set of points in convex position that is ordered in counterclockwise direction.
$P$ is called a right (resp. left) half convex point set if $Q$ is upper-right monotonic and
$R$ is lower-right monotonic (resp. $Q$ is upper-left monotonic and $R$ is lower-left monotonic).
\end{definition}

\begin{definition}\emph{(Centrally symmetric convex point set)}
Let $P$ be a set of points in convex position. $P$ is said to be centrally
symmetric with respect to the origin, if for every point $p \in P$, point $-p$ also belongs to $P$
\end{definition}
Let $p,q,r$ be three points in $P$.
\begin{lemma} \label{lemma:angle}
If $q$ and $r$ are neighbors of $p$ in an LGG on $P$, then $\angle{pqr}, \angle{prq} < \pi/2$.
\end{lemma}
\proof
Since $(p,q)$ is an edge of $G_P$, $r$ must lie outside the disk $d_{pq}$.
Thus, $\angle{prq} < \frac{\pi}{2}$. Since $(p,r)$ is also an edge in $G_P$, $q$ must lie
outside the disk $d_{pr}$. Thus, $\angle{pqr} < \frac{\pi}{2}$.
\qed

\vspace{0.1in}
Conversely, if either $\angle{pqr} \geq \frac{\pi}{2}$ or $\angle{prq} \geq \frac{\pi}{2}$, then we call
the edges $(p,q)$ and $(p,r)$ as {\it conflicting}. Two conflicting edges cannot exist simultaneously in an LGG.

\section{Lower Bound Construction}
In this section, we describe the construction of a $LGG$ with $\Omega(n^{5/4})$ edges.
The point set $P$ for this construction is a $\sqrt n \times \sqrt n$ uniform grid. First, we describe
the algorithm that constructs the $LGG$ $G_P$ on the grid point set $P$. Then, we prove the
correctness of our algorithm. Finally, we analyze the edge complexity of $G_P$.

\subsection{Construction}
Let us denote the points on the grid as $(x,y), 0 \leq x,y < \sqrt n$. The algorithm is an iterative greedy
procedure that assigns neighbors to each grid point. First, we describe the procedure
that assigns neighbors to an arbitrary point $p=(p^x,p^y)$ on the grid.
For technical reasons, we only assign neighbors to $p$ that are in the first and third quadrant w.r.t. $p$.
By applying this procedure to the grid points $(x,y), \sqrt n/3 \leq x,y < 2\sqrt n/3$ (we choose only
these grid points to avoid edge effects), we obtain our $LGG$ $G_P$.

\begin{figure}[htp]
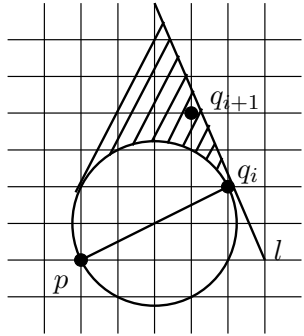

\centering
\input feasibility.pstex_t
\caption{Feasibility region for the next neighbor $q_{i+1}$}\label{fig:feasible}
\end{figure}

Now, we describe the iterative procedure that assigns neighbors to $p$ in a counter-clockwise direction.
Let $q_i$ be the current neighbor of $p$ that is assigned by the procedure
and $\theta_i$ be the angle that segment $pq_i$ makes with the positive direction of x-axis.
First, we describe how to find the next neighbor $q_{i+1}$ in the counter-clockwise direction.
Let us describe the feasibility region for $q_{i+1}$. Figure~\ref{fig:feasible} shows the points $p,q_i$,
the disk $d_{pq_i}$ and the tangent line $l$ at $q_i$. Since $(p,q_i)$ is an edge in $G_P$, $q_{i+1}$
must lie outside $d_{pq_i}$. Also, since $(p,q_{i+1})$ will be an edge in $G_P$, $\angle{pq_iq_{i+1}} < \frac{\pi}{2}$
(by Lemma~\ref{lemma:angle}). This implies that $q_{i+1}$ must lie below the tangent line $l$.
Thus the feasible region for $q_{i+1}$ is outside $d_{pq_i}$ and below $l$ (shown as the shaded region in
Figure~\ref{fig:feasible}).
We choose the next neighbor $q_{i+1}$ to be the grid point in the feasible region that is closest
(in Euclidean distance) to $q_i$ (See Figure~\ref{fig:feasible}). This greedy choice allows us
to pack as many neighbors as possible.

Now, the procedure that assigns neighbors to $p$ is as follows: The first neighbor of $p$
is set as $q_0=(p^x+s,p^y+s\cdot \tan\theta_0)$, where $s=\sqrt n/3$ and $\theta_0, 0 < \theta_0 < \pi/4$
is a small constant to be fixed later. Starting with this neighbor, we iteratively
find the next neighbor using the procedure described above. We continue assigning neighbors as long as the condition
$\theta_i \leq \pi/4$ is satisfied. Note that this procedure assigns neighbors only in the first quadrant w.r.t $p$.
Similarly, we find neighbors in the third quadrant w.r.t $p$ by starting with the initial neighbor $(p^x-s,p^y-s\cdot \tan\theta_0)$
and proceeding as long as the condition $\theta_i \leq 5\pi/4$ is satisfied.


\subsection{Correctness}
In this section, we show that the geometric graph $G_P$ constructed above is a locally gabriel graph.

{\bf Remark 1:} Observe that the above procedure that constructs $G_P$ assigns neighbors in a symmetric consistent manner, i.e.,
{\em if the procedure assigns $q_i$ as the $i$-th neighbor (in 1st quadrant) of $p$, then it would
assign $p$ as the $i$ th neighbor (in 3rd quadrant) of $q_i$, when the procedure is applied
on $q_i$.}

By Remark 1, the neighbors of $p$ in $G_P$ are exactly the grid points chosen by the procedure.

\begin{figure}[htp]
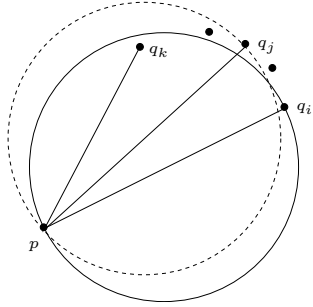

\centering
\input correctness.pstex_t
\caption{Point $p$ and its neighbors $q_i,\dots,q_j,\dots,q_k$}
\label{fig:correctness}
\end{figure}

\begin{lemma} \label{lemma:correctness}
Let $p\in P$ be any grid point and let $Q=\{q_0,q_1,\dots,q_m\}$ be the neighbors of $p$ in $G_P$ (in counter-clockwise order) in the first quadrant.
The disk $d_{pq_i}$ does not contain any neighbor of $p$ $\forall i, 0 \leq i \leq m$.
\end{lemma}
\proof
First, we show that $d_{pq_i}$ does not contain any neighbor of $p$ in the first quadrant, i.e.,
$d_{pq_i} \cap (Q \setminus \{q_i\}) = \emptyset$.
Observe that $d_{pq_i}$ does not contain $q_{i+1}$ because the iterative procedure
picks $q_{i+1}$ outside the disk $d_{pq_i}$. Also observe that $d_{pq_i}$ does not contain
$q_{i-1}$ because $\angle{pq_{i-1}q_i} < \frac{\pi}{2}$ ($q_i$ is picked below tangent line
of $d_{pq_{i-1}}$).
On the contrary, let us assume that
$d_{pq_i}$ contains some $q_j, j\neq i-1,i,i+1$. There are 2 cases:
(i) $j>i+1$ and (ii) $j<i-1$. We will prove case (i) below. Case (ii)
can be proved in a similar manner. Let us assume that $k$ is the smallest index among the neighbors
$q_j, j>i+1$ that is contained in $d_{pq_i}$. Since $q_i,q_{i+1},\dots q_{k-1},q_k$ are in counter-clockwise
convex position, all the disks $d_{pq_j}, i+1 \leq j \leq k-1$ also contains $q_k$ (see figure~\ref{fig:correctness}).
Thus, the disk $d_{pq_{k-1}}$ also contains $q_k$.
This is a contradiction since the iterative procedure picks $q_k$ outside the disk
$d_{pq_{k-1}}$.

The disk $d_{pq_i}$  does not contain any neighbor of $p$ in the third quadrant w.r.t $p$, since $d_{pq_i}$ does
not intersect the third quadrant w.r.t $p$. Thus $d_{pq_i}$ does not contain any neighbor of $p$.
\qed

{\bf Remark 2:} Observe that the grid point set $P$ is a symmetric point set and we use the same deterministic procedure to assign
neighbors to all the grid point. Hence Lemma \ref{lemma:correctness} is true for all the grid points $p\in P$.

\begin{lemma} \label{lemma:p}
Edge $(p,q_i)$ of $G_P$ satisfies the locally Gabriel condition $\forall i, 0 \leq i \leq m$.
\end{lemma}
\proof
We need to show that the disk $d_{pq_i}$, $0 \leq i \leq m$, does not contain
the neighbors of $p$ or $q_i$ in $G_P$. By Lemma \ref{lemma:correctness}, disk $d_{pq_i}$ does not
contain any neighbor of $p$.

By Remark 1, $p$ is the $i$ th neighbor (in 3rd quadrant) of $q_i$.
By Remark 2, we apply Lemma \ref{lemma:correctness} for grid point $q_i$ (instead of $p$) on the neighbors of $q_i$ in the 3rd
quadrant (instead of 1st quadrant) to show that disk $d_{q_ip}$ (which is the same as $d_{pq_i}$) does not contain any
neighbors of $q_i$.
\qed

Since the procedure assigns neighbors to $p$ in the third quadrant in exactly the same way as the first quadrant,
Lemma \ref{lemma:p} shows that edges from $p$ to its neighbors in the third quadrant also satisfy the
locally Gabriel condition.
Thus, all the edges from $p$ to neighbors of $p$ satisfies the locally
Gabriel condition. Since we use the same deterministic procedure to assign neighbors to all the grid points,
the argument for $p$ applies to all grid points $p\in P$. Hence all the edges in $G_P$ satisfy the locally Gabriel condition
proving that $G_P$ is locally Gabriel.

\subsection{Analysis}
In this section, we analyze the lower bound construction described in the previous section.
We will show that $G_P$ has $\Omega(n^{5/4})$ edges
by proving that the iterative procedure picks $\Omega(n^{1/4})$ neighbors for grid point $p$.
Let $q_0,q_1,\dots,q_m$ be the neighbors(in counter-clockwise order) of $p$ in the first quadrant.
Given the current neighbor $q_i$, the procedure picks the next neighbor $q_{i+1}$ ``close'' to $q$.
We will prove bounds on the closeness between $q_i$ and $q_{i+1}$. Using this, we show bounds on $m$.

\begin{figure}[htp]
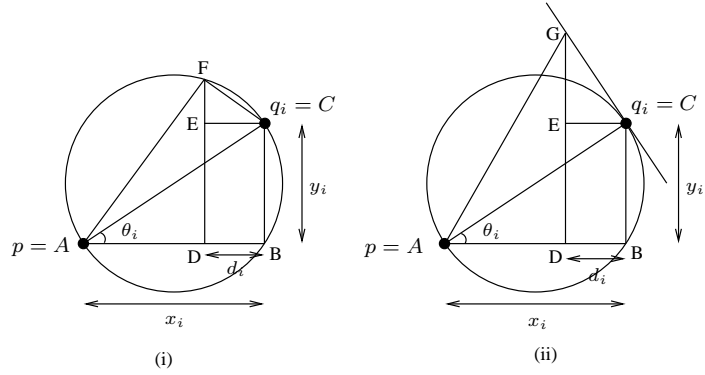

\centering
\input analysis.pstex_t
\caption{The vertical gridline that contains the next neighbor intersects (i) the diameter disk at $F$, (ii) the tangent line at $G$ }
\label{fig:analysis1}
\end{figure}

Figure~\ref{fig:analysis1} shows the points $p$ (denoted as A in the figure), the current neighbor $q_i$ (denoted as C),
the disk $d_{pq_i}$ and the tangent line $l$ at C. Let the next neighbor $q_{i+1}$ lie at a x-distance $d_i$
from the current neighbor $q_i$ ($q_{i+1}$ lies on the vertical line passing through D and E).
Let $\vert AB \vert = q_i^x-p^x = x_i$, $\vert DB \vert = q_i^x - q_{i+1}^x = d_i$ and 
$\vert CB \vert = q_i^y-p^y = y_i$ (See Figure~\ref{fig:analysis1}). First, we will
prove bounds for $d_i$ in terms of $x_i$. Let the vertical grid-line passing through $q_{i+1}$ intersect
the disk $d_{pq_i}$ at $F$ (See Figure~\ref{fig:analysis1}(i)) and the tangent line at $G$
(See Figure~\ref{fig:analysis1}(ii)). Let $\vert FE \vert =h_i$ and $\vert GE \vert =h'_i$. 
Since $\triangle AFC$ is right-angled at $F$ (see Figure~\ref{fig:analysis1}(i)), we have
\begin{align*}
{\vert AC \vert}^2 &= {\vert AF \vert}^2 + {\vert FC \vert}^2 \\
&= ({\vert AD \vert}^2 + {\vert DF \vert}^2) + ({\vert FE \vert}^2 + {\vert CE \vert}^2) \\
{(x_i \sec \theta_i)}^2&= {(x_i-d_i)}^2 + {(h_i+ x_i \tan \theta_i)}^2 + h_i^2 + d_i^2
\end{align*}
Simplifying, we get
\begin{equation}
h_i^2 + x_i \tan \theta_i \cdot h_i - d_i(x_i-d_i) = 0
\label{eq:case1}
\end{equation}
\newline
Similarly, since $\triangle ACG$ is right-angled at $C$ (see Figure~\ref{fig:analysis1}(ii)),
we have
\begin{align*}
{\vert AG \vert}^2 &= {\vert AC \vert}^2 + {\vert CG \vert}^2 \\
({\vert AD \vert}^2 + {\vert DG \vert}^2) &= {\vert AC \vert}^2 + ({\vert GE \vert}^2 + {\vert CE \vert}^2) \\
{(x_i-d_i)}^2 + {(h'_i+ x_i\tan \theta_i)}^2 &=  {(x_i \sec \theta_i)}^2 + {h'_i}^2 + d_i^2
\end{align*}
Simplifying, we get $h'_i = d_i \cot \theta_i$.

The next neighbor $q_{i+1}$ lies on the vertical gridline between $F$ and $G$. 
To ensure that a grid point exists between $F$ and $G$, we enforce a stronger 
condition that the distance between $F$ and $G$ is at least $1$,
i.e., $\vert FG \vert = h'_i-h_i>1$. Solving for $h_i$ in Equation~\ref{eq:case1}, substituting for $h_i,h'_i$,
we get
\begin{equation}
d_i \cot \theta_i - \frac{\sqrt{x_i^2\tan^2 \theta_i + 4d_i(x_i-d_i)} - x_i \tan \theta_i}{2} > 1
\label{eq:main}
\end{equation}
Simplifying this, we get the inequality
\begin{equation*}
d_i^2 + \sin^2 \theta_i > x_i \tan \theta_i \sin^2 \theta_i + d_i \sin 2\theta_i
\end{equation*}
\remove{
Since $\theta_i$ has the range $0 \leq \theta_i \leq \pi/4$, it simplifies to
\begin{equation*}
d_i^2 + \sin^2 \theta_i > x_i + d_i \sin 2\theta_i
\end{equation*}
}
By setting $d_i = c_1\sqrt{x_i}, c_1 > 1$, the above inequality is satisfied, since 
$\theta_i \leq \pi/4$ (we assign neighbors to $p$ only till $\theta_i \leq \pi/4$). 
Therefore, inequality ~\ref{eq:main} is also satisfied. This gives us a bound on $d_i$
(closeness between $q_{i+1}$ and $q_i$) in terms of $x_i$ (x-distance of $q_i$ from $p$).
\remove{
Thus, if the x-coordinate distance of current neighbor $q_1$ from $p$ is $c$,
the iterative procedure finds the next neighbor at x-coordinate distance
$c-2\sqrt{c}$ from $p$. Note that we start the iterative procedure from the initial
neighbor that is distance $t$ from $p$ and we stop when the x-coordinate of neighbor
is less than $p^x$. Let $x_i$ denote the x-coordinate distance of $q_i$ from $p$ for all $i$.
Then, we have the recurrence relation $x_{i+1}=x_i-2\sqrt{x_i},~x_0=t$. It can be easily
seen that $x_m\geq 0$ if $m=\theta(\sqrt{t})$. Since $t=\sqrt{n}/3$, we get $m=\theta(n^{1/4})$.
}

Now, we will obtain bounds on $m$, the number of neighbors assigned to $p$.
Note that the procedure assigns neighbors to $p$ as long as 
$\theta_i \leq \pi/4$, i.e., $y_m \leq x_m$.
We will now obtain bounds on $x_i$ and $y_i$. The $x_i$ are related by the
following recurrence relation
\begin{align*}
x_{i+1} &= x_i - d_i \\
&= x_i - c_1\sqrt{x_i} \\
&\geq x_i - c_1\sqrt{\frac{\sqrt{n}}{3}} \hspace{1cm} \bigl(x_i \leq \frac{\sqrt{n}}{3}\bigr)
\end{align*}
Expanding this recurrence with $x_0 = \sqrt{n}/3$ , we get
\begin{equation}
x_k \geq \frac{\sqrt{n}}{3} - \frac{k \cdot c_1n^{1/4}}{\sqrt{3}}, 0 < k \leq m
\label{eq:xk}\end{equation}
Next, we obtain bounds on $y_i$. The $y_i$ are related by the
recurrence relation $y_{i+1} = y_i + \lfloor h_i + 1 \rfloor$
(since we pick $q_{i+1}$ as the closest grid point to F).
Expanding this recurrence, we get
\begin{align}
y_k &= y_0 + \sum_0^{k-1} \lfloor h_i + 1 \rfloor \\
&\leq y_0 + k + \sum_0^{k-1} h_i
\label{eq:yi}
\end{align}
where $h_i$ is given by the solution to Equation~\ref{eq:case1} 
\begin{align*}
\sum_0^{k-1}h_i &= \frac{1}{2} \sum_0^{k-1} \sqrt{x_i^2\tan^2 \theta_i + 4d_i(x_i-d_i)} - x_i \tan \theta_i \\
&= \frac{1}{2} \sum_0^{k-1} \sqrt{x_i^2\tan^2 \theta_i + 4c_1\sqrt{x_i}(x_i-c_1\sqrt{x_i})} - x_i \tan \theta_i \\
&= \frac{1}{2} \sum_0^{k-1}x_i \tan \theta_i  \biggl(\sqrt{1 + \frac{4c_1\sqrt{x_i}(x_i-c_1\sqrt{x_i})}{x_i^2\tan^2 \theta_i}} - 1 \biggr) \\ 
&\leq \frac{1}{2} \sum_0^{k-1}x_i \tan \theta_i  \biggl(\sqrt{1 + \frac{4c_1}{\sqrt{x_i}\tan^2 \theta_i}} - 1 \biggr) \\
&\leq \frac{1}{2} \sum_0^{k-1}x_i \tan \theta_i \biggl( \Bigl(1 + \frac{2c_1}{\sqrt{x_i}\tan^2 \theta_i} \Bigr) - 1 \biggr) \\ 
&\leq \sum_0^{k-1} \frac{c_1\sqrt{x_i}}{\tan \theta_i}
\end{align*}
Since $\theta_i > \theta_0$ and $x_i \leq \sqrt{n}/3$, we have 
\begin{equation*}
\sum_0^{k-1}h_i \leq \frac{c_1 \cdot k \cdot n^{1/4}}{\sqrt{3}\tan \theta_0}
\end{equation*}
Hence, from Equation~\ref{eq:yi}, $y_k$ is given by the following
\begin{equation}
y_k \leq \frac{\tan \theta_0 \cdot \sqrt{n}}{3} + \frac{c_1 \cdot k \cdot n^{1/4}}{\sqrt{3}\tan \theta_0} + k
\label{eq:yk}\end{equation}
Setting $c_1=1.01, \theta_0=1.74 \times 10^{-3}$, it can be verified analytically in Equation~\ref{eq:xk} and Equation~\ref{eq:yk} 
that $y_k \leq x_k$ for all $0 \leq k \leq 10^{-4} n^{1/4}$.
Thus, $y_m \leq x_m$ for $m = 10^{-4} n^{1/4}$. The number of neighbors of $p$ is at least $10^{-4} n^{1/4}$.
The edge complexity of $G_P$ is therefore $\Omega(n\cdot n^{1/4})=\Omega(n^{5/4})$.

\section{Convex Point Sets}


In this section, we show edge complexity for LGG on various classes of convex point sets.
First, we show exact bounds for half convex point sets. Then, we show asymptotic tight
linear bounds for special subclasses of convex point sets. Finally, we show $O(n\log n)$
bounds for arbitrary convex point sets.

\subsection{Exact Bound for Half Convex Point Sets}

First, let us consider the special case when $P$ is a monotonic convex point set.
Wlog, let us assume that $P$ is of the upper-right type.
\begin{lemma} \label{lemma:monotonic}
Let $P=\{p_1,p_2,\dots,p_n\}$ be a upper-right monotonic convex point set and let
$G_P$ be any locally gabriel graph on $P$. $p_1, p_n$ has atmost 1 neighbor in $G_P$
and hence $G_P$ has atmost $n-1$ edges.
\end{lemma}
\proof
We show that the first point $p_1$ has atmost one neighbor. Let if possible, $p_i$ and $p_j$
be neighbors of $p_1, j > i$. $p_1, p_i, p_j$ are in monotonic convex position. Thus $\angle{p_1p_ip_j} \geq 90^\circ$.
Since $p_i$ and $p_j$ are neighbors of $p_1$, $\angle{p_1p_ip_j} < 90^\circ$ (by Lemma~\ref{lemma:angle}).
Hence a contradiction. By a similar argument, we can also show that $p_n$ has atmost 1 neighbor in $G_P$.

Removing $p_1$ from $P$ and applying induction on the remaining points, we see that $G_P$ has
atmost $n-1$ edges.
\qed

Next we consider the special case when $P$ is a half convex point set. Wlog, let us assume that $P$ is
a right half convex point set.

\begin{lemma} \label{lemma:half}
Let $P=Q\cup R$ be a right half convex point set with $n$ points, where $Q$ is upper-right
monotonic and $R$ is lower-right monotonic. Let $G_P$ be any locally gabriel graph on $P$.
$G_P$ has atmost $2n-3$ edges.
\end{lemma}
\proof
Let $p$ be the point with maximum x-coordinate (rightmost point) in $P$.
$Q \cup \{p\}$ is upper-right monotonic and $R \cup \{p\}$ is lower-right monotonic.
By Lemma~\ref{lemma:monotonic}, $p$ has degree atmost two(atmost one neighbor
in $Q$ and one in $R$). Removing $p$ from $P$ and applying induction on the
remaining points, we get $P(n)\leq P(n-1)+2;P(2)=1$.
This gives $P(n) \leq 2n-3$.
\qed

The above bounds are tight, i.e., it is easy to construct locally gabriel graphs
for monotonic and half convex sets that match the above bounds. For any monotonic
convex sets, construct a path (of length $n-1$) connecting all the vertices.

\begin{figure}[htbp]
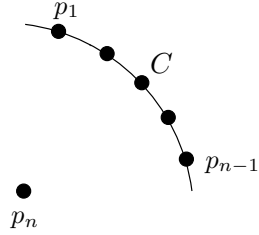

\centering
\input circle.pstex_t
\caption{Portion of circle C centered at $p_n$ and points $p_1,\dots,p_{n-1}$ placed equidistant on $C$}
\label{fig:circle}
\end{figure}

For right half convex sets, we can achieve the exact bound using the following construction: \\
Let $C$ be a circle with center at $p_n$. We place points $p_1, p_2, \dots, p_{n-1}$ equidistant
along the first quadrant of $C$ (See Figure~\ref{fig:circle}). The point set constructed is right half convex.
The edges of $G_P$ are defined as follows: \\ \\
(i) Add edges $(p_i, p_{i+1}), 1 \leq i \leq n-2$. This forms a path of length $n-2$. \\ \\
(ii) Add edges $(p_n, p_i), 1 \leq i \leq n-1$. This forms a star of size $n-1$. \\ \\
It can be verified that these edges satisfy the locally gabriel condition. Thus, the edge
complexity of $G_P$ is $2n-3$.

\subsection{Tight Linear Bounds for Various Subclasses}
In this section, we prove asymptotic tight linear bounds for some special subclasses
of convex point sets.

\subsubsection{Points on a Circle}
First, we consider the special case of $n$ points lying on a circle.

\begin{figure}[htbp]
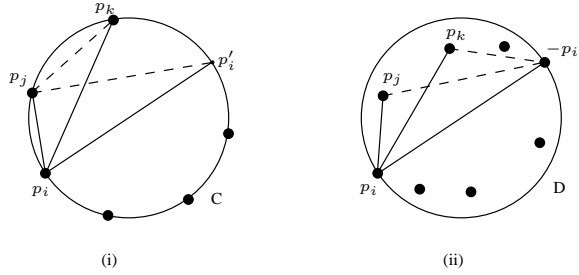

\centering
\input subclass_convex.pstex_t
\caption{(i) Points on a circle $C$ (ii) Centrally symmetric point set with diameter pair $p_i,-p_i$}
\label{fig:subclass}
\end{figure}

\begin{lemma}
Let $C$ be any circle and $P=\{p_1,p_2,\dots,p_n\}$ be $n$ points that lie on $C$.
Let $G_P$ be any locally gabriel graph on $P$. $G_P$ has atmost $n$ edges
\end{lemma}
\proof
Let $p_i$ be any point in $P$ and $p'_i$ be the point on $C$ that is diametrically
opposite to $p_i$. The diameter $p_ip'_i$ divides the circle $C$ into two halves.
We claim that $p_i$ has atmost 1 neighbor in each half. Let, if possible, $p_i$ have
two neighbors $p_j$ and $p_k$ in the same half (see Figure~\ref{fig:subclass}(i)). We can see that
$\angle{p_ip_jp'_i} = 90^\circ$. Since $p_i, p_j, p_k, p'_i$ are in convex position,
we have $\angle{p_ip_jp_k} >  \angle{p_ip_jp'_i}$. Thus, $\angle{p_ip_jp_k} > 90^\circ$
But, since $(p_i,p_k)$ is an edge, $\angle{p_ip_jp_k} < 90^\circ$ (by Lemma~\ref{lemma:angle}).
Hence a contradiction.

Since, each point $p_i \in P$ has atmost 2 neighbors (atmost one in each half),
the edge complexity of $G_P$ is atmost $n$.
\qed

This bound is exact, since we can always construct a $G_P$ with $n$ edges.

\subsubsection{Centrally symmetric convex point set}
Next, we consider the case of $P$ being in centrally symmetric convex position.
We prove that any locally gabriel graph on $P$ has atmost $2n-3$ edges. Our proof is an adaptation of~\cite{AM02},
where it was proved that the unit distance graph on centrally symmetric convex point sets has
atmost $2n-3$ edges.

\begin{lemma}
Let $P=\{p_1,-p_1,p_2,-p_2\dots,p_{n/2},-p_{n/2}\}$ be $n$ points in centrally symmetric convex position.
Let $G_P$ be any locally gabriel graph on $P$. $G_P$ has atmost $2n-3$ edges
\end{lemma}

\proof
In~\cite{AM02}, it is shown that the diameter pair (pair that is furthest apart)
in any centrally symmetric convex point set must be of the form $(p_m, -p_m)$,
for some $m$. Let $(p_i, -p_i)$ be the diameter pair in $P$.

If $(p_i, -p_i)$ is an edge in $G_P$, we can show (by a similar argument as below)
that $p_i, -p_i$ has atmost 1 neighbors in $P$. Thus, $G_P$ would have atmost $n-1$ edges.
Therefore, let us assume that $(p_i, -p_i)$ is not an edge in $G_P$.

Consider the closed disk $D$ with $p_i$ and $-p_i$ as diameter. Since $P$ is centrally
symmetric, all the points in $P$ must lie in $D$. The diameter $p_i,-p_i$ divides the disk $D$ into two halves.
We claim that $p_i$ has atmost 1 neighbor in each half. Let, if possible, $p_i$ have
two neighbors $p_j$ and $p_k$ in the same half (see Figure~\ref{fig:subclass}(ii)). Since, $p_j$ lies in $D$,
$\angle{p_ip_j-p_i} \geq 90^\circ$.  Also, since $p_i, p_j, p_k, -p_i$ are in convex position,
$\angle{p_ip_jp_k} >  \angle{p_ip_j-p_i}$. Thus, $\angle{p_ip_jp_k} > 90^\circ$.
Since $(p_i,p_k)$ is an edge, $\angle{p_ip_jp_k} < 90^\circ$ (by Lemma~\ref{lemma:angle}).
Hence a contradiction.

$p_i$ has atmost 2 neighbors in $G_P$. By the same argument, $-p_i$ also has atmost
2 neighbors. Removing $p_i$ and $-p_i$ from $P$ and recursing on the remaining point set
(which is also centrally symmetric), we have $P(n)\leq P(n-2)+4; P(2)=1$.
This gives $P(n) \leq 2n-3$.
\qed

We can achieve an almost tight lower bound using the following construction: \\
Let $P$ be a set of $n$ points defined by $P = \{ (-1,i) \cup (1,i), -n/4 \leq i < n/4 \}$.
$P$ consists of equally spaced integer gridpoints on the vertical lines $x=-1$ and $x=1$
($n/2$ points in each line). It is easy to see that $P$ is centrally symmetric about the origin.
The edges of $G_P$ are defined as follows: \\ \\
(i) Add $n-4$ edges of the form $\bigl( (-1,i), (-1,i+2) \bigr)$ and $\bigl( (1,i) , (1,i+2) \bigr)$
for all $-\frac{n}{4} \leq i < \frac{n}{4}-2$. \\ \\
(ii) Add $n-4$ edges of the form $\bigl( (-1,i), (1,i+1) \bigr)$ and $\bigl( (-1,i), (1,i-1) \bigr)$
for all $-\frac{n}{4}-1 \leq i < \frac{n}{4}-1$. \\ \\
It can be easily verified that these edges satisfy the locally gabriel condition. Thus, the edge complexity of $G_P$ is $2n-8$.

\subsection{Bounds for Convex Point Sets}

In this subsection, we consider an arbitrary convex point set $P$. We prove that the edge complexity
of any LGG on $P$ is $O(n\log n)$. The proof is a straightforward extension
of the argument given in \cite{BP01}, which proved that the unit distance graph on convex point sets has
$O(n\log n)$ edges.

\remove{
We use the following lemma of \cite{BP01} in our proof.
\begin{lemma}[\cite{BP01}] \label{lemma:BP01}
Let $P$ be a set of $n$ points in convex position and let $G_P$
be a geometric graph on $P$. Let $P$ be partitioned into a right half
convex set $Q$ and a left half convex set $R$ using any antipodal pair
and let $G_Q, G_R$ be the vertex-induced subgraphs of $G_P$. If $G_Q$
and $G_R$ have $O(n)$ edges, then $G_P$ has $O(n\log n)$ edges
\end{lemma}

The above lemma is proved using a simple, recursive argument. We use Lemma~\ref{lemma:half}
and Lemma~\ref{lemma:BP01} to prove the following:
}

\begin{lemma}
Let $P$ be a set of $n$ points in convex point set and let $G_P$ be
any locally gabriel graph on $P$. $G_P$ has $O(n\log n)$ edges.
\end{lemma}
\remove{
\proof
Let $P$ be partitioned into a right half convex set $Q$ and a left half convex set $R$
using an antipodal pair and let $G_Q, G_R$ be the vertex-induced subgraphs of $G_P$.
Since $Q$ and $R$ are half convex, $G_Q, G_R$ have $O(n)$ edges (by Lemma~\ref{lemma:half}).
Applying Lemma~\ref{lemma:BP01}, we can see that $G_P$ has $O(n\log n)$ edges.
\qed
}

\proof
We use the clever recursive method given in~\cite{BP01}. We will describe the method
briefly, for sake of completeness. Refer to~\cite{BP01} for details.
We can partition $P$ into $Q$ and $R$ using the topmost and bottommost point of $P$ (antipodal pair).
Note that $Q$ is left half convex and $R$ is right half convex.
In fact, we can perform a partition using any of the antipodal pairs, such that the two parts are
half convex sets (for an appropriate reference axis).
The basic idea behind the recursive method in~\cite{BP01} is to use the above fact to divide $P$ using
two such partitions such that we have two subproblems of size atmost $3n/4$ and the edges at this level
of recursion are edges within the four half convex sets. The number of such edges is $O(n)$ using Lemma~\ref{lemma:half}.
The edge complexity of $G_P$ is thus $O(n \log n)$
\qed

\remove{
We use 2 such partitions to divide $P$ into $Q_1, Q_2, R_1, R_2$, where $Q_1 \cup Q_2$ and $R_1 \cup R_2$
are half convex (by 1st partition) and $Q_1 \cup R_2$ and $Q_2 \cup R_1$ are half convex (by 2nd partition)
and $\vert Q_1 \vert, \vert Q_2 \vert \geq n/4$ (See Figure ???). Let $E(P)$ denote the number of edges in $G_P$
and $E(P_1,P_2)$ denotes the number of edges of $G_P$ that go between $P_1$ and $P_2$.
We get the following recurrence relation
\begin{align*}
E(P) &= E(Q_1, R_1) + E(Q_2, R_2) + E(Q_1 \cup Q_2) + E(R_1 \cup R_2) + E(Q_1 \cup R_2) + E(Q_2 \cup R_1)
\end{align*}
}

\remove{
\proof
Let $P=Q \cup R$, where $Q$ is left half convex and $R$ is right half convex.
The edges in $G_P$ can be partitioned into two sets based on the endpoints:
(i) $E_1$ is the set of edges with both endpoints in $Q$ or $R$ and (ii) $E_2$ is
the set of edges going between $Q$ and $R$.

By Lemma~\ref{lemma:half}, $E_1$ has $O(n)$ edges. We bound the size of $E_2$ using
the following lemma of Furedi~\cite{F}.

\begin{lemma}[Furedi] \label{lemma:Furedi}
Let $Q$ and $R$ be disjoint arcs of a convex polygon with $n$ vertices.
The number of unit distance pairs between $Q$ and $R$ is $O(n \log n)$.
\end{lemma}

The above lemma is based on showing a particular forbidden subgraph in the unit distance graph.
It can be shown that this subgraph is also forbidden for locally gabriel
graph on convex sets. Thus, $E_2$ has atmost $O(n \log n)$ edges.
\qed
}

For convex point sets, the best known lower bound is $2n-3$.

\section{Independent Sets}
In this section, we show that any $LGG$ on any $n$ point set contains an independent set of size at least $\Omega(\sqrt{n}\log n)$.

We first show an elementary argument that constructs an independent set of size at least $\frac{\sqrt{n}}{2}$ in a $n$ point set.
A set of points ordered by their abscissa is called a monotonic sequence if the ordinates of the points are either
monotonically non-increasing or monotonically non-decreasing.
\begin{lemma} \label{lemh}
Let $G_P$ be any $LGG$ on a monotonic sequence $P$ with $n$ points. $G_P$ has an independent set of size at least $\frac{n}{2}$.
\end{lemma}
\proof
Let us denote the first and the last vertices of the monotonic sequence $P$ as terminal vertices.
We show that in any $LGG$ on $P$, a terminal vertex has degree at most one.
On the contrary let us assume that a terminal vertex $v$ is incident to vertices $v_1$ and $v_2$ and the vertices appear in the
sequence as $v,v_1$ and $v_2$.
An axis parallel rectangle with $vv_2$ as diagonal will contain $v_1$ inside or on the boundary of it. It
implies that edges $(v,v_2)$ and $(v,v_1)$ conflict with each other. Thus, $v$ has at most one edge incident to it.
Now, add the terminal vertex to the independent set and remove it along with its neighbor (if it exists) from the sequence. In each iteration
at most two vertices are removed and one vertex is added to the independent set. Thus, the independent set has size at least $\frac{n}{2}$.
\qed

Erdos and Szekeres \cite{Erdos} showed that a set of $n$ points will have a monotonic sequence of size at least $\sqrt{n}$.
One such sequence can be computed in $O(n \log n)$ time by an algorithm proposed by Hunt and Szymanski \cite{Hunt}.
By Lemma~\ref{lemh}, any induced $LGG$ on this monotonic sequence has an independent set of size at least $\frac{\sqrt{n}}{2}$.

Now, we show that any $LGG$ on any point set with $n$ points contains an independent set of size at least $\Omega(\sqrt{n}\log n)$.
In a graph $G = (V,E)$ for any $u \in V$, let us define $N(u) = \{ v \ |\ (u,v) \in E\}$. A graph is said to have sparse neighborhood if for any $u \in V$, the
chromatic number of the subgraph induced over vertices $\{u\} \cup N(u)$ is a constant.
We show that any $LGG$ with $n$ vertices will have an independent set of size $\Omega(\sqrt{n}\log n)$
by using Theorem~\ref{aln} where the sparse neighborhood property of $LGGs$ (shown in the Lemma~\ref{4clr}) is applied.
\begin{theorem}\label{aln}
(Alon~\cite{alon}) Let $G = (V, E)$ be a graph on n vertices with average degree $t \ge 1$ in which for
every vertex $v \in V$ the induced subgraph on the set of all neighbors of $v$ is r-colorable. Then, the
independence number of $G$ is at least $\frac{c}{\log (r+1)}\frac{n}{t} \log t$, for some absolute positive constant c.
\end{theorem}
\begin{lemma} \label{4clr}
Let $G_P$ be any $LGG$ on any point set $P$ and $u$ be an arbitrary vertex in $G$. The induced subgraph over the vertices $\{u\} \cup \{N(u)\}$ is 4-colorable.
\end{lemma}
\proof
Let vertex $u$ be adjacent to $v_1,v_2,\ldots,v_k$. Let us consider the induced subgraph over these vertices. We show
that any vertex say $v_1$ has at most one incident edge on either side of the line passing through $u$ and $v_1$.
On the contrary let us assume that there are two vertices $v_2$ and $v_3$ adjacent
to $v_1$ on the same side of line $\overline{uv_1}$. Let us analyze all the possible cases.
\begin{figure}[!h]
\centering
\input{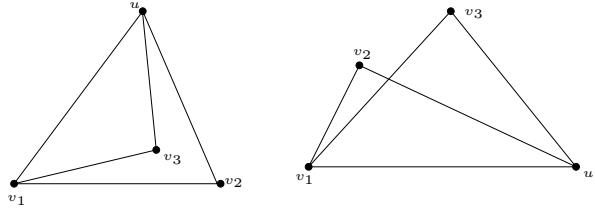}
\caption{Possible placement of neighborhood in $LGG$}
\label{fig7}
\end{figure}
\begin{itemize}
\item All the four vertices ($u,v_1,v_2$ and $v_3$) cannot be collinear otherwise at least two vertices (say $v_1$ and $v_2$ w.l.o.g.) lie on the same side of $u$ and
the edges $(u,v_1)$ and $(u,v_2)$ would conflict with each other.
\item Let us consider the case when three vertices are collinear. It can be trivially verified that $v_1, v_2$ and $v_3$ cannot be collinear due to
$LGG$ constraints. Similarly $u, v_1$ and $v_2$ (or $v_3$) also cannot be collinear due to $LGG$ constraints.
If $v_2,v_3$ and $u$ are collinear then $u$ must lie in between $v_2$ and $v_3$. It contradicts with the assumption that $v_2$ and $v_3$ lie on the same side of $\overline{uv_1}$.
\item Let us consider the case when convex hull of these four vertices is a triangle and another vertex lies inside this triangle as shown in Figure~\ref{fig7}(a). Since it is assumed that $v_2$ and $v_3$ lie on the
same side of $\overline{uv_1}$, $u$ and $v_1$ must be the vertices of this triangle. Let us assume that vertex $v_3$ lies inside $\triangle uv_1v_2$. Since
$(u,v_1),(u,v_2)$ and $(u,v_3)$ do not conflict with each other, both $\angle uv_3v_1$ and $\angle uv_3v_2$ should be less than $\frac{\pi}{2}$, which is
not possible in this configuration.
\item The last case is when all the vertices are in convex position and form a quadrilateral.
Lets assume w.l.o.g. that $uv_1v_2v_3$ is a convex quadrilateral as shown in Figure~\ref{fig7}(b). By Lemma~\ref{lemma:angle}, $\angle uv_1v_2 < \frac{\pi}{2}$ (due to edges $uv_1$ and $uv_2$),
$\angle v_1v_2v_3 < \frac{\pi}{2}$ (due to edges $v_1v_3$ and $v_1v_2$), $\angle v_2v_3u < \frac{\pi}{2}$ (due to edges $uv_2$ and $uv_3$), $\angle v_3uv_1 < \frac{\pi}{2}$ (due to edges $v_1u$ and $v_1v_3$) and .
But in a quadrilateral at least one of the internal angle should be greater than or equal to $\frac{\pi}{2}$. Hence, it leads to a contradiction.

\end{itemize}
Hence any vertex $v_i \in N(u)$ has at most two neighbors apart from $u$ in the induced subgraph on neighborhood of $u$.
Thus, the degree of any vertex $v_i$ for $1 \le i \le k$ is at most 3. Therefore, this induced subgraph is 4-colorable.
\qed
\begin{theorem}
 Let $G_P$ be any $LGG$ on a $n$ point set. $G_P$ has an independent set of size $\Omega(\sqrt{n}\log n)$.
\end{theorem}
\proof
 Since an $LGG$ has a maximum of $O(n^\frac{3}{2})$ edges~\cite{KL10}, the average degree of a vertex is $O(\sqrt{n})$. Substituting $t=O(\sqrt{n})$ and $r=4$ in Theorem~\ref{aln}, the desired bound is obtained.
\qed 
\section*{Conclusion}
In this paper, we have shown improved bounds on the maximum edge complexity
of locally gabriel graphs. There is still a gap between our lower bound of $\Omega(n^{5/4})$ 
and the best known upper bound of $O(n^{3/2})$. It is an interesting problem to narrow this gap.
We have shown tight linear bounds for various subclasses of convex pointsets. But, for a general 
convex point sets, the best lower bound on edge complexity of locally gabriel graphs
is $2n-3$, while the upper bound is $O(n \log n)$. Can one obtain tight bounds?
Finally, we have shown that any LGG on any $n$ pointset has an independent set of size $\Omega(\sqrt{n}\log n)$.
There is no known non-trivial upper bound. It is an interesting problem to improve upon these bounds.

\begin{spacing}{0.9}
\bibliographystyle{abbrv}
\bibliography{lgg}
\end{spacing}
\end{document}